\documentclass{article}
\usepackage{cite}
\usepackage{spconf,amsmath,amssymb,graphicx,hyperref}

\title{B-GRPO: Unsupervised Speech Emotion Recognition based on Batched-Group Relative Policy Optimization}
\name{Yingying Gao$^{1,2}$, Shilei Zhang$^{1,2}$, Runyan Yang$^{1,2}$, Zihao Cui$^{1,2}$, Junlan Feng$^{1,2}$
 }
\address{$^1$JIUTIAN Research, China Mobile, Beijing, China\\ 
$^2$The State Key Laboratory of Multimedia Information Processing, Peking University, Beijing, China
 }
\begin{document}
%\ninept
%
\maketitle
\begin{abstract}
Unsupervised speech emotion recognition (SER) focuses on addressing the problem of data sparsity and annotation bias of emotional speech. Reinforcement learning (RL) is a promising method which enhances the performance through rule-based or model-based verification functions rather than human annotations. We treat the sample selection during the learning process as a long-term procedure and whether to select a sample as the action to make policy, thus achieving the application of RL to measure sample quality in SER. We propose a modified Group Relative Policy Optimization (GRPO) to adapt it to classification problems, which takes the samples in a batch as a group and uses the average reward of these samples as the baseline to calculate the advantage. And rather than using a verifiable reward function as in GRPO, we put forward self-reward functions and teacher-reward functions to encourage the model to produce high-confidence outputs. Experiments indicate that the proposed method improves the performance of baseline without RL by 19.8\%.
\end{abstract}
\section{Introduction}
\label{sec:intro}

Speech emotion recognition (SER) faces the persistent challenge of capturing natural and spontaneous emotional speech, and the annotation process is time-consuming and involves individual perceptual bias. 

Recently, unsupervised or self-supervised learning methods have shown promise to alleviate this issue, which extract speech features from large scale unlabeled general speech and build emotion classifiers upon these feature extractors. One way to constraint the unsupervised training is contrastive learning \cite{cl3}, which augments the input via perturbations and attempts to capture the invariant consistent features between the augmented samples. The other way is self-constraint \cite{sl1}, by which the model is trained to reconstruct the input or predict the masked tokens of the input. The reconstruction process enables the model to extract more abstract and contextual features. DINO \cite{c1} is also a self-constraint method but based on distillation, guiding the learning of the student network via a supposed teacher network, which has the same architecture but different parameters with the student network, and updates with an exponential moving average (EMA) of the student parameters. Above all, these constraint methods enable the unsupervised models to capture the most essential or generic features, which might contribute to eliminating the individual deviations in emotion expression and perception.

Reinforcement learning (RL) is also an effective way to improve model performance when the annotated training data is limited. However, the application of RL for emotion recognition still faces challenges. First, RL focuses on maximizing the cumulative expected reward for long-term goals, while SER is a classification problem without a cumulative reward. Secondly, the results in generative models such as Generative Pre-trained Transformers (GPTs) \cite{gpt} have randomness due to random sampling in generating process, so that an input query can generate multiple responses. In the recently proposed Group Relative Policy Optimization (GRPO) \cite{c2} algorithm, these responses have been used as a group to calculate the relative advantage, replacing the requirement of a value function in Proximal Policy Optimization (PPO) \cite{c3} and significantly reducing memory and computation consumption. However, the prediction of SER remains fixed without multiple candidates generated for a same input, which means GRPO can not be adopted directly. 

Therefore, we treat the training of SER model as a long-term process, and whether to take an advantage of one sample at each step as the actions to make policy, thus achieving the selection of the unlabeled samples by RL. Additionally, we propose a modified GRPO to adapt the application in SER, which takes a batch of samples as a group and calculate the normalized rewards for the responses within each batch as the advantage function, termed as Batched-GRPO (B-GRPO). Experiments have shown that batched normalization achieve better results. Moreover, instead of using a verifiable reward function as in GRPO, we propose \textit{self-reward} function and \textit{teacher-reward} function to pick out useful samples without annotations. The self-reward function is determined by the likelihood probability, and the teacher-reward function decides whether a sample is available through additional teacher model. After comparison, self-reward is superior. We employ likelihood probability as the self-reward function, since a larger likelihood probability means a larger confidence of the predicted results, and as the learning progresses, the model will learn more accurate parameter estimates.

\section{Related Work}
\label{sec:format}

\begin{figure*}[htbp]
    \centerline{\includegraphics[width=\textwidth]{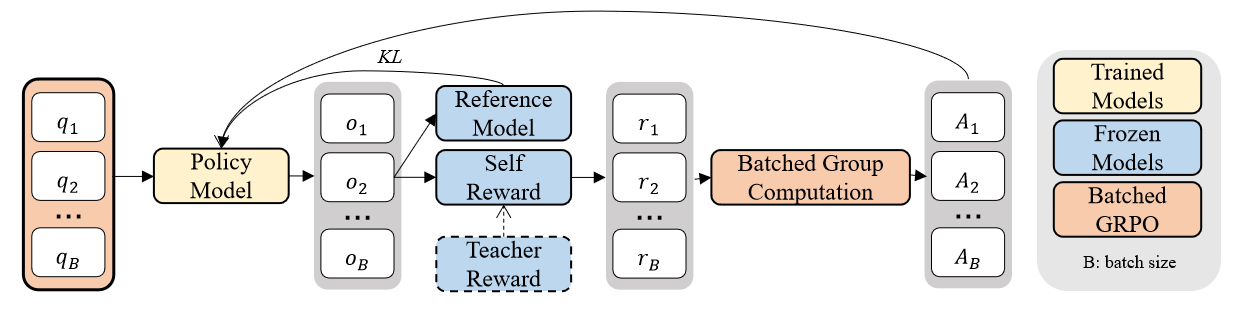}}
    \caption{The framework of the proposed B-GRPO.}
    \label{fig1}
\end{figure*}
%\vspace{-0.5cm}
\textbf{Unsupervised Speech Emotion Recognition:} The methods for unsupervised SER can be roughly divided into two categories: data generation and self-supervised pre-training. \cite{c6} decouples speech signals into semantic and paralinguistic components and converts the paralinguistic components to generate new data for low-resourced emotion categories. \cite{c7} proposes corpus aware emotional CycleGAN to synthesize target-aware samples to augment source datasets. There is similar work in unsupervised transfer learning \cite{c9} to solve the domain mismatch problem in SER. The other type of unsupervise SER methods are representation learning based on pre-training. \cite{c11} proposes Emotion2vec based on masking prediction and knowledge distillation to supply powerful representations for SER. \cite{c12} adopts a denoising autoencoder (DAE) and an adversarial autoencoder (AAE) to extract the features from LibriSpeech for model pre-training. \cite{c13} proposes contrastive predictive coding (CPC) which is able to learn salient representations from unlabeled datasets and improve emotion recognition performance.

\textbf{Reinforcement Learning:} \cite{c4} and \cite{c5} also propose unsupervised RL methods that improves reasoning performance by using the own confidence of the model as a reward, \cite{c4} uses negative entropy while \cite{c5} adopts the average KL divergence between a uniform distribution over the vocabulary and the next-token distribution as the self-certainty metric. They show potential of self-constraint, but none of them are proposed for classification problems. R1-Omni \cite{c14} presents the first application of RL on SER with verifiable reward (RLVR) to an Omni-multimodal LLM. This work is able to use GRPO directly since the SER depends on reasoning process which can generate multiple candidates to form a group, but the computation of the reward function depends on the ground truth which is not applicable to unlabeled data. \cite{c15} proposes a self-play paradigm that a single model learns to propose tasks that maximize its own learning progress and find that tasks which are too difficult or too simple are less valuable, similar with our findings that samples with flatter likelihood probability distributions are less reliable.

\section{Method}
\label{sec:pagestyle}

We propose a modified GRPO algorithm, Batched-GRPO, for the application in unsupervised emotion recognition. The core idea of GRPO emphasizes relative rather than absolute optimization of policy, which leverages the average reward of multiple sampled outputs of the same query as the baseline, obviating the need for additional value function approximation as in PPO. We instead take the samples in a batch as a group and use the average reward of these samples as the baseline to calculate the advantage, as shown in the orange boxes in Fig.~\ref{fig1}. In this context, the advantage function no longer measures the superiority of different responses to the same question, but the quality of each sample within the same batch. The policy model is the SER model. The action refers to whether to use this sample in the update of policy model. Rather than using a verifiable reward function as in GRPO, we propose self-reward functions and teacher-reward functions, in which the former is determined by the maximum likelihood probability of the prediction, and the later is decided through the comparison between the prediction of the current policy model and the outcomes from external teacher models.

The proposed self-reward functions are shown in equation (1) and (2):
\begin{equation}
    r1=\begin{cases}
        C & \text{ if } \max_N{p_{p}(n|q_i)}>\delta \\
        0 & \text{ else } 
      \end{cases}
\end{equation}
%\vspace{-0.5cm}
\begin{equation}
    r2=\max_N{p_{p}(n|q_i)} 
\end{equation}
where $p_{p}(n|q_i)$ is the likelihood probability predicted by policy model on emotion $n$ for input sample $q_i$, $\max_N$ is the operation of taking the maximum value among $N$ emotional categories. $\delta$ is a threshold. $C$ is a constant greater than 0 as the reward score. Since $\max_N{p_{p}(n|q_i)}$ is also greater than 0, we set it as the reward score in $r2$. Both $r1$ and $r2$ adopt the own assessment of the model as feedback, without relying on external supervision or labeled target responses. Higher likelihood probability indicates greater confidence of the prediction and peaked distributions at emotion categories. These rewards encourages the model to produce high-confidence outputs.

Besides self-reward based on the outputs of policy model, we also introduce external models as teacher models to make the results more confident. The teacher models are not updated during training. The teacher-reward functions include:
\begin{equation}
    r3=\begin{cases}
        C & \text{ if } \text{argmax}_N{p_{p}(n|q_i)}==\text{argmax}_N{p_{t}(n|q_i)} \\
        0 & \text{ else } 
      \end{cases}
\end{equation}
\begin{equation}
    r4=\begin{cases}
        C & \text{ if } r1 \wedge r3 \\
        0 & \text{ else } 
      \end{cases}
\end{equation}
\begin{equation}
    r5=\begin{cases}
        C & \text{ if } D_{KL}(P_{t}(n|q_i)\parallel P_{p}(n|q_i)) <\theta  \\
        0 & \text{ else } 
      \end{cases}
\end{equation}
in which $r3$ represents that if the predicted category of the policy model is consistent with the result of the teacher model.  $r4$ means that the conditions of $r1$ and $r3$ to receive rewards are satisfied simultaneously. $r5$ employs K-L divergence of the probability distributions between teacher model and policy model to measure the consistency of the predicted results. $\theta$ is a threshold. The teacher-reward comprises both the feedback of policy model and external teacher model to supply more confident results but the performance is inferior to self-reward in this work.

To further eliminate the negative impact of defective samples, we modify the advantage function and exclude the parts less than 0: 
%\begin{equation}
%    A_{i} =\frac{r_{i}-\bar{r_{i} }  }{\hat{r_{i}} } 
%\end{equation}
\begin{equation}
    \hat{A_{i}}=\begin{cases}
        A_{i}=\frac{r_{i}-\bar{r_{i} }  }{\hat{r_{i}} }  & \text{ if } A_{i}>0 \\
        0 & \text{ else } 
       \end{cases} 
\end{equation}
where $r_{i}$ is the reward of the $i^{th}$ sample, $\bar{r_{i}}$ is the mean value of the rewards within the batch and $\hat{r_{i}}$ is the standard error.

In the context of SER, the sequence length is not needed. For each batch with size B, the GRPO loss becomes:
\begin{equation}
    \begin{aligned}
    & \Im_{B-GRPO} = \frac{1}{B}   \sum_{i  = 1}^{B}\left \{ \min \right [ \frac{\pi_{\theta}\left( o_i\mid q_i\right)}{\pi_{{\theta }_{old}}\left(o_i\mid q_i \right)} \hat{A_i}, \\
    & \text{clip}\left (\frac{\pi_{\theta}\left ( o_i\mid q_i\right )}{\pi_{{\theta }_{old}}\left (o_i\mid q_i \right )},1-\varepsilon, 1+\varepsilon \right )\hat{A_i} \left ]-\beta \mathbb{D}_{KL}\left [ \pi _{\theta } \parallel \pi _{ref}\right ]  \right \}
    \end{aligned}
\end{equation}
\begin{equation}
    \mathbb{D}_{KL}\left [ \pi _{\theta } \parallel \pi _{ref}\right ]=\frac{\pi_{ref}\left( o_i\mid q_i\right)}{\pi_{\theta}\left(o_i\mid q_i \right)}-\log{\frac{\pi_{ref}\left( o_i\mid q_i\right)}{\pi_{\theta}\left(o_i\mid q_i \right)}
} -1 
\end{equation}

\section{Experiments}
\label{sec:typestyle}

\textbf{Setup:} We implement the experiments via Emobox \cite{c16} toolkit, which takes pre-trained speech encoders as feature extractors. Only the last transformer layer is adopted as the feature for each frame and the average of the frame features is taken as the utterance feature. We take the SenseVoice \cite{c17} model as the feature extractor for policy model. The feature extractor is not updated. 
The policy model comprises two linear hidden layers and a nonlinear layer with a ReLU activation function between them. 
% and compare other feature extractors such as Emotion2vec-plus-large \cite{c11} and Whisper-large-v3 \cite{c18}. 
The teacher reward model has the same structure with policy model but takes other models as the feature extractor, including Emotion2vec-plus-large \cite{c11}, Emotion2vec-base \cite{c11} and Whisper-large-v3 \cite{c18}.
% We also test other feature extractors for teacher reward, including Emotion2vec-plus-large \cite{c11}, Emotion2vec-base \cite{c11} and Whisper-large-v3. 
The learning rate is 1e-4 and the hidden size is 128. The batch size is 32. We also test other batch sizes since it relates with group size, and find that 32 or 64 achieves the best balanced performance among all datasets. %The performance changes more significantly when the group size decreases, but not obviously when the group size increases.

\textbf{Datasets:} We select five datasets in Emobox, including IEMOCAP \cite{c19}, CASIA \cite{c20}, CAFE \cite{c21}, MELD \cite{c22} and M3ED \cite{c23}, spanning 3 languages, in which 2 in English, 2 in Mandarin and 1 in French. All the annotations are mapped to six emotion categories: Neutral, Angry, Surprise, Sad, Happy and Fear.

\vspace{-0.2cm}
\subsection{Results}
We first compare the proposed B-GRPO with the baseline without RL and the other unsupervised learning method DINO \cite{c1}. The baseline is trained by half of labeled corpus for supervised learning for 100 epochs. This step is crucial for equipping the model with preliminary SER capability before proceeding to unsupervised training. Then the other half dataset is employed by B-GRPO and DINO without labels and trained for another 100 epochs. The results listed in Table~\ref{tab1} shows that B-GRPO increases the F1 scores of baseline on the five datasets by 2.2\%, 48.0\%, 16.3\%, 21.3\%, and 11.5\%, respectively, and improves by an average of 10.3\% compared to DINO.

\vspace{-0.5cm}
\begin{table}[htbp]
    %\vspace{-0.5cm}
    \caption{The performance (macro F1 score (F1\%)) of B-GRPO and the comparison with baselines}
    \begin{center}
        \setlength{\tabcolsep}{0.05cm}
    \begin{tabular}{|l|c|c|c|c|c|}
    \hline
    & IEOMCAP & CASIA & CAFE & MELD & M3ED \\
    \hline
    Baseline & 67.7	& 25.0	& 44.7	& 25.3	& 28.8\\
    \hline
    DINO & 69.2	& 28.5	& 51.0	& 26.6	& 30.8 \\
    \hline
    %200 epochs half labeled & & & & &\\
    Same epochs & 68.6 & 29.5	& 48.7	& 27.3	& 29.7 \\
    \hline
    %200 epochs full labeled & & & & &\\
    Full labeled & 69.2	& \textbf{57.2}	& 50.3	& 28.3	& 31.5\\
    \hline
    B-GRPO & \textbf{69.2}	& 37.0	& \textbf{52.0}	& \textbf{30.7}	& \textbf{32.1} \\
    \hline
    \end{tabular}
    \label{tab1}
    \end{center}
\end{table}
\vspace{-0.5cm}

In order to eliminate the impact of the increase of training epochs, we compare the results of training the same epochs without B-GRPO, as presented in the third row in Table~\ref{tab1}, B-GRPO still has an advantage ranged from 0.9\% to 25.4\% across five corpora. The fourth row is the results that is trained with all labeled data in each corpus for 200 epoch without B-GRPO. The comparison with B-GRPO indicates that B-GRPO has the potential to surpass or approach the performance with more supervised data.

\vspace{-0.2cm}
\subsection{Reward Function}

We compare different reward functions with a reward value of 1. We test other reward value and penalty value (2 and -1), but the difference is not significant. We take the SenseVoice as the feature extractor of policy model. For the teacher-reward, we compare different models as the feature extractor, 
%including Emotion2vec-plus-large \cite{c11}, Emotion2vec-base \cite{c11} and Whisper-large-v3 \cite{c18}, 
as shown in Table~\ref{tab2}, some of which achieves the best score on individual corpus, but in terms of performing better on all datasets, self-reward is optimal.
In the end, we conduct other experiments via the self-reward $r1$, which estimates whether the maximum likelihood probability is greater than the threshold of 0.5.
\vspace{-0.5cm}
\begin{table}[htbp]
    \caption{The comparison of different reward functions (F1\%)}
    \begin{center}
    \begin{tabular}{|l|c|c|c|c|c|}
    \hline
    & IEOMCAP & CASIA & CAFE & MELD & M3ED \\
    \hline
    \multicolumn{6}{|c|}{self-reward \ SenseVoice} \\
    \hline
    $r1$ & 69.2	& 37.0	& 52.0	& 30.7	& \textbf{32.1}\\
    \hline
    $r2$ & 69.3	& 36.3	& 51.9	& \textbf{31.0}	& 31.7 \\
    \hline
    \multicolumn{6}{|c|}{teacher-reward \ Emotion2vec-plus-large} \\
    \hline
    $r3$ & 69.1	& 33.7	& 51.5	& 30.4	& 32.0 \\
    \hline
    $r4$ & 69.3	& 36.3	& \textbf{52.6}	& 30.3	& 31.8 \\
    \hline
    $r5$ & 69.8	& 29.8	& 50.7	& 30.5	& 31.4 \\
    \hline
    \multicolumn{6}{|c|}{teacher-reward \ Emotion2vec-base} \\
    \hline
    $r3$ & \textbf{70.0}	& 34.4	& 50.1	& 30.0	& 31.0 \\
    \hline
    $r4$ & 69.6	& \textbf{37.5}	& 51.6	& 30.3	& 31.8 \\
    \hline
    \multicolumn{6}{|c|}{teacher-reward \ Whisper-large-v3} \\
    \hline
    $r3$ & 24.5	& 13.4	& 8.8	& 10.7	& 9.2 \\
    \hline
    $r4$ & 69.2	& 35.8	& 51.1	& 30.6	& 31.9 \\
    \hline
    \end{tabular}
    \label{tab2}
    \end{center}
\end{table}
\vspace{-1.0cm}
%\vspace{-0.9cm}
\subsection{Advantage Function}
We also test the results without the advantage function or without positive advantage only (shown in Table~\ref{tab3}), both of which are inferior than the proposed positive advantage.
\vspace{-0.5cm}
\begin{table}[htbp]
    \caption{The effect of advantage function (F1\%)}
    \begin{center}
     \setlength{\tabcolsep}{0.05cm}
    \begin{tabular}{|l|c|c|c|c|c|}
    \hline
    & IEOMCAP & CASIA & CAFE & MELD & M3ED \\
    \hline
    $\hat{A_i}>0$ & 69.2	& \textbf{37.0}	& \textbf{52.0}	& 30.7	& \textbf{32.1}\\
    \hline
    w/o\ $\hat{A_i}>0$ & 69.2	& 35.4	& 51.8	& \textbf{30.8}	& \textbf{32.1} \\
    \hline
    w/o\ $\hat{A_i}$ & \textbf{69.7}	& 36.2	& 51.8	& 29.7	& 31.3 \\
    \hline
    \end{tabular}
    \label{tab3}
    \end{center}
\end{table}
\vspace{-1.0cm}
%\vspace{-0.9cm}

\subsection{Policy Model}
We compare the performance of policy models with different feature extractors. The policy models are first trained with all the labeled data in each corpus for 100 epochs as the baseline, then trained by B-GRPO for another 100 epochs. Table~\ref{tab4}  indicates that B-GRPO is able to select effective samples either labeled or not. Among them, Whispe-large-v3 with B-GRPO has the largest improvement, while Emotion2vec-pluse-large with B-GRPO has relatively limited improvement.
\vspace{-0.5cm}
\begin{table}[htbp]
    \caption{The performance (F1\%) of policy models with different feature extractors}
    \begin{center}
        \setlength{\tabcolsep}{0.05cm}
    \begin{tabular}{|r|c|c|c|c|c|}
    \hline
    & IEOMCAP & CASIA & CAFE & MELD & M3ED \\
    \hline
    Sensevoice & 69.4	& 53.5	& 47.6	& 27.7	& 31.1\\
    \hline
    w B-GRPO & 70.1	& 59.2	& 53.3	& 31.8	& 33.9 \\
    \hline
    Emotion2vec & 70.4	& 68.3	& 83.5	& 28.1	& 25.5 \\
    \hline
    w B-GRPO & 69.9	& 70.2	& 83.6	& 29.9	& 28.5 \\
    \hline
    Whisper & 70.6	& 47.9	& 55.1	& 31.7	& 33.6 \\
    \hline
    w B-GRPO & 72.0	& 55.2	& 61.9	& 36.7	& 36.2 \\
    \hline
    \end{tabular}
    \label{tab4}
    \end{center}
\end{table}
\vspace{-1.0cm}

%\subsection{Group Size}

%Table~\ref{tab5} presents that the medium group size (32 and 64) achieves best balanced performance among all datasets. The performance changes more significantly when the group size decreases, but not obviously when the group size increases.
%\vspace{-0.5cm}
%\begin{table}[htbp]
 %   \caption{The performance (F1\%) with different group sizes}
 %   \begin{center}
 %       \setlength{\tabcolsep}{0.05cm}
 %   \begin{tabular}{|r|c|c|c|c|c|}
  %  \hline
  %  Group size & IEOMCAP & CASIA & CAFE & MELD & M3ED \\
 %   \hline
  %  8 & \textbf{71.1}	& 59.0	& \textbf{54.2}	& 30.1	& 31.8\\
  %  \hline
 %   16 & 70.6	& 58.1	& 53.0	& 31.5	& 33.1 \\
 %   \hline
 %   32 & 70.1	& \textbf{59.2}	& 53.3	& 31.8	& \textbf{33.9} \\
  %  \hline
 %   64 & 69.8	& 58.8	& 53.3	& \textbf{32.0}	& 33.6 \\
 %   \hline
 %   128 & 69.7	& 58.4	& 52.5	& 31.8	& 33.5 \\
 %   \hline
  %  256 & 69.6	& 58.6	& 52.1	& 31.9	& 33.4 \\
  %  \hline
  %  \end{tabular}
  %  \label{tab5}
  %  \end{center}
    %\vspace{-0.5cm}
%\end{table}
%\vspace{-0.5cm}

\subsection{Dataset}
In this section, we compare the data adopted by B-GRPO from the same corpus or external corpus to demonstrate the ability of data selection or data augmentation. The results in Table~\ref{tab6}  indicates that B-GRPO is able to select valid data from original corpus or external corpus, even the speech content or language changes, and introducing external corpus is less helpful than using the same corpus, but the two can be stacked in the future.%Therefore, rather than saying that the model is selecting data, it is more accurate to say that it is choosing parameters to make the model more compatible with the current dataset.
%external data attracts the attention of the model to the original data.
%The reason is that our reward is related to the likelihood probability. The higher the likelihood probability, the higher the fitness of the model to the data. The addition of external data attracts the attention of the model to the original data. Therefore B-GRPO is not dedicated to augmenting the original corpus but updating model parameters that best fit the current data. 
\vspace{-0.5cm}
\begin{table}[htbp]
    \caption{The performance (F1\%) of B-GRPO adopted different data, the left of the arrow ($\rightarrow$) is the dataset implemented B-GRPO, while the right is the data trained the baseline model. The performance is evaluated on the right corpus.}
    \begin{center}
    \begin{tabular}{|l|c|c|c|}
    \hline
    & Baseline & Same  & External \\
    & & Corpus & Corpus\\
    \hline
    SAVEE$\rightarrow$MELD  & 27.7	& 31.8	& 29.5\\
    \hline
    JL$\rightarrow$IEMOCAP & 69.4	& 70.1	& 69.6 \\
    \hline
    M3ED$\rightarrow$CASIA & 53.5	& 59.2	& 47.0 \\
    \hline
    CASIA$\rightarrow$M3ED & 31.1	& 33.9	& 32.3 \\
    \hline
    CASIA$\rightarrow$CAFE & 47.6	& 53.3	& 51.5 \\
    \hline
    CAFE$\rightarrow$CASIA & 53.5	& 59.2	& 58.2 \\
    \hline
    Librispeech$\rightarrow$MELD & 27.7	& 31.8	& 29.9 \\
    %train-clean-100 & & & \\
    \hline
    \end{tabular}
    \label{tab6}
    \end{center}
\end{table}
\vspace{-0.5cm}

\vspace{-0.2cm}
\section{Conclusion}
\label{sec:majhead}

We propose an unsupervised emotion recognition method based on RL. The sample selection during the learning process is regarded as a long-term process, and different strategies are conducted on different samples. By calculating the relative advantages of samples within each batch, we have achieved selecting valid unlabeled or labeled samples. Compared with another common unsupervised learning algorithm DINO, the performance is improved by an average of 10.3\% on five open source datasets, and increased by 19.8\% relative to the baseline without the RL stage. Experiments reveal that the success benefits from the self-reward that enhances the accuracy of model estimates by sharpening the likelihood probability distribution.

\end{document}